\documentclass[referee,fleqn,usenatbib]{mnras}

\usepackage[utf8]{inputenc}

\usepackage{orcidlink} 

\usepackage{newtxtext,newtxmath}
\usepackage{afterpage} % in your preamble

\usepackage[T1]{fontenc}

\DeclareRobustCommand{\VAN}[3]{#2}
\let\VANthebibliography\thebibliography
\def\thebibliography{\DeclareRobustCommand{\VAN}[3]{##3}\VANthebibliography}

\usepackage{graphicx}	% Including figure files
\usepackage{amsmath}	% Advanced maths commands

\title[Constraining Exoplanetary Magnetism]{Stellar Magnetic Storm Induced Magnetospheric Polarity Reversals: Distinguishing between Unmagnetised and Magnetised Exoplanets}

\author[Gupta et al.]{
Sakshi Gupta\orcidlink{0000-0002-8920-1495},$^{1,2,3}$\thanks{E-mail: sg20rs057@iiserkol.ac.in}
Souvik Roy\orcidlink{0000-0002-0471-4591},$^{1}$
Dibyendu Nandy\orcidlink{0000-0001-5205-2302},$^{1,2}$\thanks{E-mail: dnandi@iiserkol.ac.in}
\\
$^{1}$Center of Excellence in Space Sciences India, Indian Institute of Science Education and Research Kolkata, Mohanpur 741246, India\\
$^{2}$Department of Physical Sciences, Indian Institute of Science Education and Research Kolkata, Mohanpur 741246, India\\
$^{3}$Center for High Energy Systems and Sciences, Defence Research and Development Organization, Hyderabad, 500058, India
}

\usepackage{fix-cm}
\pubyear{\the\year{}}

\begin{document}
\label{firstpage}
\pagerange{\pageref{firstpage}--\pageref{lastpage}}
\maketitle

\begin{abstract}
Exoplanetary and planetary environments are forced by stellar activity which manifest through variable radiation, particle and magnetic fluxes, stellar winds, flares and magnetic storms known as coronal mass ejections (CMEs). Recent studies have shown that (exo)planets with intrinsic magnetic fields and magnetospheres respond differently to this stellar forcing compared to planets which lack an intrinsic magnetism; this is borne out by observations in solar system planets. However, detailed investigations to uncover the subtle ways in which stellar magnetic storms impact exoplanets are still at a nascent stage. Here we utilize 3D magnetohydrodynamic simulations to investigate the impact of stellar CMEs on Earth-like planets with different magnetic fields. Our results show that planetary atmospheric mass loss rates are dependent on the relative orientation of stellar wind and planetary magnetic fields, with significantly higher losses when the CME and planetary magnetic fields are oppositely oriented -- favoring enhanced magnetic reconnections. In contrast, for unmagnetised planets, the mass loss rate do not strongly depend on stellar magnetic field orientation. More significantly, we find that stellar CME induced polarity reversals can distinguish between planets with and without intrinsic magnetism. In unmagnetised or weakly magnetised (exo)planets, the polarity of the externally imposed magnetosphere are prone to global polarity reversals forced by stellar magnetised storms. Our analysis of the magnetotail current density dynamics during polarity reversals aligns with observations of Venus.  This distinction in magnetospheric response provides a new paradigm to differentiate between (exo)planets with or without significant (intrinsic) magnetic fields.
\end{abstract}

% \begin{keywords}
% Stellar winds; Magnetohydrodynamical simulations; Planetary magnetospheres; Exoplanet atmosphere; Exoplanet atmospheric evolution; Star-planet interactions; Stellar magnetic fields
% \end{keywords}

%%%%%%%%%%%%%%%%%%%%%%%%%%%%%%%%%%%%%%%%%%%%%%%%%%

%%%%%%%%%%%%%%%%% BODY OF PAPER %%%%%%%%%%%%%%%%%%

\section{Introduction} \label{sec:intro}

The variable magnetism of stars, and the evolution of this activity influences (exo)planetary space environments in myriad ways that have consequences for planetary space weather and habitability \citep{NANDY2007891,Gallet_2017,Mozos_2019}. The most energetic magnetic phenomena in stars are stellar coronal mass ejections (CMEs) -- energetic events characterized by the expulsion of large amounts of magnetised plasma from stellar corona, a phenomena widely studied and observed in the solar system \citep{Hundhausen1999, Chen_2011b}. These events -- often associated with solar-stellar flares and active region -- are critical in shaping space weather conditions around (exo)planets. While initially observed in our solar system, observational and stellar spectroscopic evidence reveals that CMEs are ubiquitous across various stellar types. Observations from space-based observatories like Kepler, TESS and Chandra X-ray have revealed frequent stellar flares and coronal mass ejection events in young active and cool dwarf stars \citet{Vida_2016, odert_2017, Argiroffi2019, 2020_Gunther, Raetz_2024}. These stellar events, analogous to solar CMEs, can be even more energetic in red dwarf stars exhibiting flares much more powerful than those observed on the Sun \citep{Kowalski_2013}. When these energetic plasma ejections propagate through the interplanetary medium and interact with planetary magnetospheres, they are referred to as Interplanetary Coronal Mass Ejections (ICMEs) \citep{Gosling_1993, Schwenn_2006}. The magnetosphere of an (exo)planet acts as a dynamic shield against the solar-stellar wind and ICMEs, regulating the transfer of energy, mass, and momentum from the stellar wind to the planetary environment \citep{Russell2001}.

Understanding how stellar wind interacts with (exo)planetary magnetic fields during strong events such as CMEs is crucial for determining the evolution of (exo)planetary atmospheres and by extension, assessing their potential habitability \citep{2003_Lammer, 2014_khodachenko, Cohen_2020, NANDY2023106081, Hazra_2024}. During stellar CMEs, particle density in the associated shock front, velocity, and the strength of the embedded magnetic field in the stellar wind increases. Thus, when a CME interacts with a planetary atmosphere, it induces significant disturbances that impact atmospheric dynamics \citep{2017_cherenkov}. \cite{Alvarado_2016} discusses stellar wind conditions in the estimated boundaries of the habitable zones (HZ) of exoplanet-host stars. For Earth, the magnetosphere's response to ICME impacts can lead to geomagnetic storms characterized by enhanced auroral activity, increased ionospheric currents, and disturbances in the radiation belts \citep{Dungey_1961, Gonzalez_1994, Gonzalez_1999}. Studies have shown that when the Interplanetary Magnetic Field (IMF) is oriented southward, it is more likely to reconnect with Earth's northward-pointing magnetic field, leading to more intense geomagnetic storms \citep{Burton1975,Gonzalez_1999}. This process, known as magnetic reconnection, is a fundamental plasma physics phenomenon that facilitates energy transfer from the solar wind into the Earth's magnetosphere, driving geomagnetic activity \citep{Sonnerup_1974,paschmann1979magnetic, Burkhart_1990}. These geomagnetic storms have far-reaching consequences, like disruptions in satellite operations, communication systems, power grids, as well as increased radiation exposure for astronauts and high-altitude flights \citep{Baker_1998, Pulkkinen_2007, Baruah_2024}. 
 
While significant progress has been made in understanding the impact of ICMEs on the Earth’s magnetosphere, detailed investigations in the context of (exo)planets are just beginning. In a recent work, \cite{Hazra_2025} explore the interaction of stellar CMEs in a close-in hot Jupiter-like system and find significant consequences. However, the response of (exo)planetary environments-spanning Unmagnetised, weakly magnetised, and strongly magnetised regimes-remain poorly characterized. Unmagnetised planets, such as Mars and Venus, interact with ICMEs in fundamentally different ways than magnetised planets like Earth. 

An outstanding challenge in (exo)planetary studies lies in distinguishing planets with and without intrinsic magnetic fields. 

Intrinsic magnetospheres arise from a planet's internal magnetic field, providing a dynamically stable magnetic shield against solar-stellar wind interactions (such as in Earth). In contrast, imposed magnetosphere forms by the interaction of the stellar wind with the planetary ionosphere in the absence of an intrinsic magnetic field (such as in Mars). The absence of a global magnetic field for these planets means that the ionosphere and atmosphere are directly exposed to stellar wind and energetic particles from stellar flares and ICMEs leading to significant atmospheric erosion \citep{Vignes_2000, Dubinin, Ramstad, basak_nandy_2021}. This atmospheric loss, driven by ion pickup and sputtering processes, has profound implications for the long-term evolution of planetary atmospheres \citep{luhmann1992solar, jakosky2001atmospheric, Lammer2008}. 

Understanding the difference between imposed and intrinsic magnetospheres is essential for analyzing (exo)planetary responses to varying stellar wind conditions. This differentiation is crucial for future space missions aimed at characterizing exoplanetary magnetic environments and assessing atmospheric retention. Analyzing the diverse magnetospheric responses can help identify optimal observational strategies for detection of planetary magnetospheres.

The relative strength of planetary to stellar wind magnetic field plays a crucial role in determining the planetary dynamics and atmospheric mass loss \citep{Gupta_2023}. Although temporary distortions and local reconnection events may occur in intrinsic magnetospheres – like that of Earth’s – the global magnetic field topology maintains its overall structure during ICME impacts \citep{MCPHERRON2008303,Zong_2022}. However, as observed for Venus and Mars, imposed magnetospheres exhibit strong variations in response to the changes in stellar wind magnetic field configurations
\citep{dubinin2013interaction, Futaana2017, Xu_2020}. These distinct responses have been observed through both in-situ measurements \citep{Slavin_1991, TROTIGNON1993189, RUSSELL20061482} and numerical simulations \citep{PREUSSE2007589,WayM_J, basak_nandy_2021}. This reconfiguration of the magnetosphere can serve as a signature for characterizing exoplanetary magnetic environments. The magnetotail lobes and current sheet dynamics also provide crucial insights into magnetospheric dynamics forced by stellar magnetism. Current density evolution in magnetotail regions is governed by the drift of charged particles along magnetic field lines, with significant variations observed during different stellar wind and ICME interactions \citep{Di_Braccio, wang_2023}. 

Previous observational studies have demonstrated diminished heavy-ion flux in near Venusian magnetotail during polarity reversal events, attributing these phenomena to magnetic detachment of plasma tails from Venusian ionospheres \citep{Vech_2016}. The magnitude of current density is intrinsically linked to the available charge carriers and the complex magnetic reconnection processes occurring during stellar wind (exo)planet interactions. Understanding these current density dynamics becomes crucial for comprehending energy transfer mechanisms, plasma interactions, and the overall response of (exo)planetary magnetospheres to varying stellar wind conditions. Atmospheric escape is another manifestation of star-planet interactions in (exo)planetary environments; e.g., it has been demonstrated that the efficiency of atmospheric loss is strongly influenced by the strength and orientations of stellar and planetary magnetic fields \citep{Gupta_2023}.

Direct observations of (exo)planetary magnetic fields have not yet been achieved, although some techniques such as Lyman $\alpha$ and and radio emissions offer promise for indirect detection through star-planet magnetic interactions \citep{Zarka1998, Zarka2007, Griessmeier2007, Hallinan2008, Shaikhislamov_2016, Kislyakova_2019}. Radio emissions have been observed in several solar system planets, providing insights into their magnetic environments \citep{Masters2012,Kurth2016,Lamy2017, Connerney2018}. Radio emissions arise from energy dissipation due to magnetic reconnection events during stellar wind-planetary magnetic field interactions. Nevertheless, direct radio detections from exoplanets remain elusive \citep{Bastian2000, Smith2009, Murphy2015, Lynch2018, Turner2021}. This creates a gap in our understanding of how planetary magnetic fields can be constrained through remote observations. In this context, magnetohydrodynamic (MHD) simulations provide a useful tool and guidance for exploring these complex star-planet interaction processes \citep{Das_2019,vidotto_2020, basak_nandy_2021, refId0, Roy_2023, Gupta_2023}. In this study, we employ 3D MHD simulations to investigate the effects of different IMF and ICME orientations on (exo)planetary dynamics, atmospheric escape rates and magnetotail current density structures for both magnetised and unmagnetised (exo)planets. We limit our study to a far out, Earth-like planetary system with varying magnetism, keeping in mind the perspective of habitable conditions, and future possibilities of detection. 

Based on our theoretical analysis, we establish a basis for distinguishing between exoplanets with and without intrinsic magnetic fields. The subsequent sections of this paper are structured as follows: Section 2 provides a comprehensive description of the theoretical framework and numerical setup utilized in our study. Section 3 details our findings, followed by concluding remarks in Section 4.

% This is a simple template for authors to write new MNRAS papers.
% See \texttt{mnras\_sample.tex} for a more complex example, and \texttt{mnras\_guide.tex}
% for a full user guide.

% All papers should start with an Introduction section, which sets the work
% in context, cites relevant earlier studies in the field by \citet{Fournier1901},
% and describes the problem the authors aim to solve \citep[e.g.][]{vanDijk1902}.
% Multiple citations can be joined in a simple way like \citet{deLaguarde1903, delaGuarde1904}.

\section{MODEL DESCRIPTION} \label{sec:style}
We adapt the 3D MHD STORM Interaction Module (CESSI-STORMI) developed by \cite{Roy_2023} to simulate the impact of different orientations of stellar wind and ICMEs on (exo)planetary magnetospheres with different intrinsic magnetic field strengths. The interactions are governed by the adiabatic equation of state and the following set of resistive MHD equations:

\begin{equation}
  \partial_{t} \rho+\nabla \cdot(\rho \vec{v})=0  
\end{equation}

\begin{equation}
   \partial_{t} \vec{v}+(\vec{v} \cdot \nabla) \vec{v}+\frac{1}{4 \pi \rho} \vec{B} \times(\nabla \times \vec{B})+\frac{1}{\rho} \nabla P=\vec{g}
\end{equation}

\begin{equation}
   \partial_{t} E+\nabla \cdot[(E+P) \vec{v}-\vec{B}(\vec{v} \cdot \vec{B})+(\eta \cdot \vec{J}) \times \vec{B}]=\rho \vec{v} \cdot \vec{g}
\end{equation}

\begin{equation}
   \partial_{t} \vec{B}+\nabla \times(\vec{B} \times \vec{v})+\nabla \times(\eta \cdot \vec{J})=0
\end{equation}

The symbols $\rho$, $\vec{v}$, $\vec{B}$, $\vec{J}$, $P$, $E$, and $\vec{g}$ denote density, velocity, magnetic field, current density, pressure, total energy density and gravitational acceleration due to the planet, respectively. We consider a finite and isotropic magnetic diffusivity $\eta$, as the causal mechanism for non-ideal processes such as magnetic reconnections. The expression for total energy density is given by

\begin{equation}
    E = \frac{P}{\gamma-1} + \frac{\rho v^2}{2} + \frac{B^2}{8\pi}
\end{equation}
for an ideal gas equation of state.

%\subsection{Grid setup}
The computational domain extends from [-145, 205 ]R$_p$ in the $x$-direction, [-100, 100]R$_p$ in the $y$-direction and [-145, 145]R$_p$ in the $z$-direction, where R$_p$ is the 
planetary radius, with the planet located at the origin (0,0,0) of the cartesian box. We use a non-uniform grid to ensure computational efficiency while attaining the desired resolution in specific regions of interest. The region extending from -5 ${R_p}$ to 5 ${R_p}$ is resolved by 40 grids in all three directions. In the $x$-direction, outermost regions extending from -145 ${R_p}$ to -100 ${R_p}$, and from 100 ${R_p}$ to 205 ${R_p}$, 1 grid corresponds to 3 ${R_p}$. Moving inward, in the regions between -100 ${R_p}$ to -50 ${R_p}$ and between 50 ${R_p}$ to 100 ${R_p}$, 1 grid corresponds to 2 ${R_p}$. In the regions extending from -50 ${R_p}$ to -25 ${R_p}$ and 25 ${R_p}$ to 50 ${R_p}$, 1 grid represents ${R_p}$. Closer to the planet from -25 ${R_p}$ to -10 ${R_p}$ and from 10 ${R_p}$ to 25 ${R_p}$, 2 grids represent ${R_p}$. Further refinement occurs between -10 ${R_p}$ to -5 ${R_p}$ and 5 ${R_p}$ to 10 ${R_p}$, where 3 grids represent ${R_p}$. The same progressive grid refinement is applied in both the y and z-directions, ensuring greater resolution near the planetary vicinity and coarser resolution farther away.

% We perform a set of simulations as shown in Fig (\ref{IMF_ICME}) by varying the stellar wind magnetic field orientations, namely northward stellar wind followed by southward ICME (nIMF - sICME), southward stellar wind followed by northward ICME (sIMF - nICME), northward stellar wind followed by northward ICME (nIMF - nICME) and southward stellar wind followed by southward ICME (sIMF - sICME). We also vary the strength of planetary magnetosphere from unmagnetised to $0.1  {B}_{{e}}, {B}_{{e}}, \thinspace $and$ \thinspace 2 {B}_{{e}}$ and perform simulations with all possible combinations. Here ${B}_{{e}}$ = $3.1\times 10^4$ nT denotes the surface equatorial magnetic field strength of Earth's dipolar field. 

\begin{table*}
        \centering
        \caption{Values of physical parameters used in the simulations and their respective notations.}
        \label{table:1}
        \begin{tabular}{lcr} 
            \hline
            \hline
            Physical quantity & Notation & Value used\\ [2ex] \hline
            Density in ambient medium  & $\rho_{amb}$ & 1.5 $\times 10^{-23}$  $\rm{g\thinspace cm^{-3}}$ \\ 
            Pressure in ambient medium & P$_{amb}$ & 2.49 $\times 10^{-11}$ $\rm{ dyne\thinspace cm^{-2}}$ \\
            Density of stellar wind in nominal conditions & $\rho_{sw (N)}$ & 4 $\rho_{amb}$\\
            Density of stellar wind during ICME & $\rho_{sw (ICME)}$ & 12.89 $\rho_{amb}$\\
            Velocity of stellar wind in nominal conditions & v$_{sw (N)}$ & 4 $\times 10^{7}$ $\rm{ cm\thinspace s^{-1}}$\\
            Velocity of stellar wind during ICME & v$_{sw (ICME)}$ & 10 $\times 10^{7}$ $\rm{ cm\thinspace s^{-1}}$\\
            Adiabatic index & $\gamma$ & 5/3\\
            Planetary mass & M$_{pl}$ & 5.97 $\times 10^{27}$ \rm{g}\\
            Planetary radius & R$_{pl}$ & 6.37 $\times 10^8$ \rm{cm}\\
            Intrinsic planetary magnetic field & B$_{pl}$ & 0 - 2B$_{{e}}$ $\thinspace^*$\\
            Stellar wind magnetic field magnitude in nominal conditions & B$_{sw (N)}$ & 10 nT \\
            Stellar wind magnetic field magnitude at the center of CME flux rope & B$_{0 cme}$ & 38 nT\\
            Magnetic diffusivity & $\eta$ & 10$^{13}$ $\rm{ cm^{2}\thinspace s^{-1}}$\\
            Magnetospheric tilt angle & $\theta_{pl}$ & 11$^\circ$\\
         \hline
        \end{tabular}

*\textit{The symbol B$_e$ represents the magnetic field for the case of the Earth's dipole}.
\end{table*}

To study the effect of stellar magnetic field orientation on planetary atmospheric loss, we approximate the atmosphere as a conducting plasma surrounding the planet. While this may overestimate the mass loss, this is a standard practice because of it is less intensive computationally; moreover, we are interested in the qualitative behavior rather then exact quantitative numbers. The density profile in the vicinity of the planet is defined by

\begin{equation*}
\rho_{{pl}} = 10^6 \rho_{{atm}}  \qquad r\leq {R_p} ,
\end{equation*}
\begin{equation}
\begin{split}
\rho_{{atm}} (r) =  \rho_{{pl}} + \frac{(\rho_{{amb}} - \rho_{{pl}})}{2} \Big[{\tanh}\Big\{ 9\Big(\frac{r}{R_P} - 2\Big)\Big\}&+1 \Big] \\
{R_P} \leq r \leq {3R_p} \thinspace ,
\end{split}
\end{equation}
where ${\rho_{{pl}}}$, ${\rho_{{amb}}}$ and $r$ represent the densities of the planet, ambient medium and radial distance from the origin respectively. For details on the choice of this atmospheric profile, refer to \cite{basak_nandy_2021}. The atmospheric distribution of pressure is calculated assuming a hydrostatic
equilibrium condition:
\begin{equation}
    \frac{{dP}}{{dr}} = -\rho_{{atm}} (r) g(r)
\end{equation}
Here, acceleration due to gravity is given by $g(r) = -\frac{G M_{{pl}}}{r^2}$ where $M_{{pl}}$ is the planetary mass. The pressure inside the planet’s core is kept constant and is equal to its value on the planetary surface. The density and pressure in the region ${r > 3 R_{{pl}}}$ is initialized to be equal to that in the ambient medium (Table \ref{table:1}).

In the model setup up the planetary dipolar magnetic field is first forced from the left boundary of our computational box with the nominal stellar wind for up to 120 minutes to establish the wind-forced, (dynamic) steady state magnetosphere \citep{Das_2019, basak_nandy_2021, Gupta_2023}. After the IMF, an ICME shock and sheath followed by an idealised ICME flux rope are injected into the computational domain. 
We model the flux ropes using the force-free nonlinear flux rope model developed by Gold and Hoyle \citep{Gold_Hoyle_1960, Hu_2014}. Following the ICME transit, we again introduce the IMF forcing on the planetary magnetosphere to simulate the relaxation of the magnetospheric system in the aftermath of the storm. In our modelled Gold-Hoyle flux rope, the twist is radially uniform and the helicity is positive. The magnetic field components in curvilinear cylindrical coordinates ($x = r cos \phi$, $x = r sin \phi$, z = z)
are:
\begin{equation*}
    B_r = 0
\end{equation*}
\begin{equation*}
    B_\phi =  \frac{Tr}{1+T^2r^2} B_0 
\end{equation*}
\begin{equation*}
    B_z =  \frac{1}{1+T^2r^2} B_0 
\end{equation*}

\begin{figure*}
    \centering
    \includegraphics[width=\textwidth]{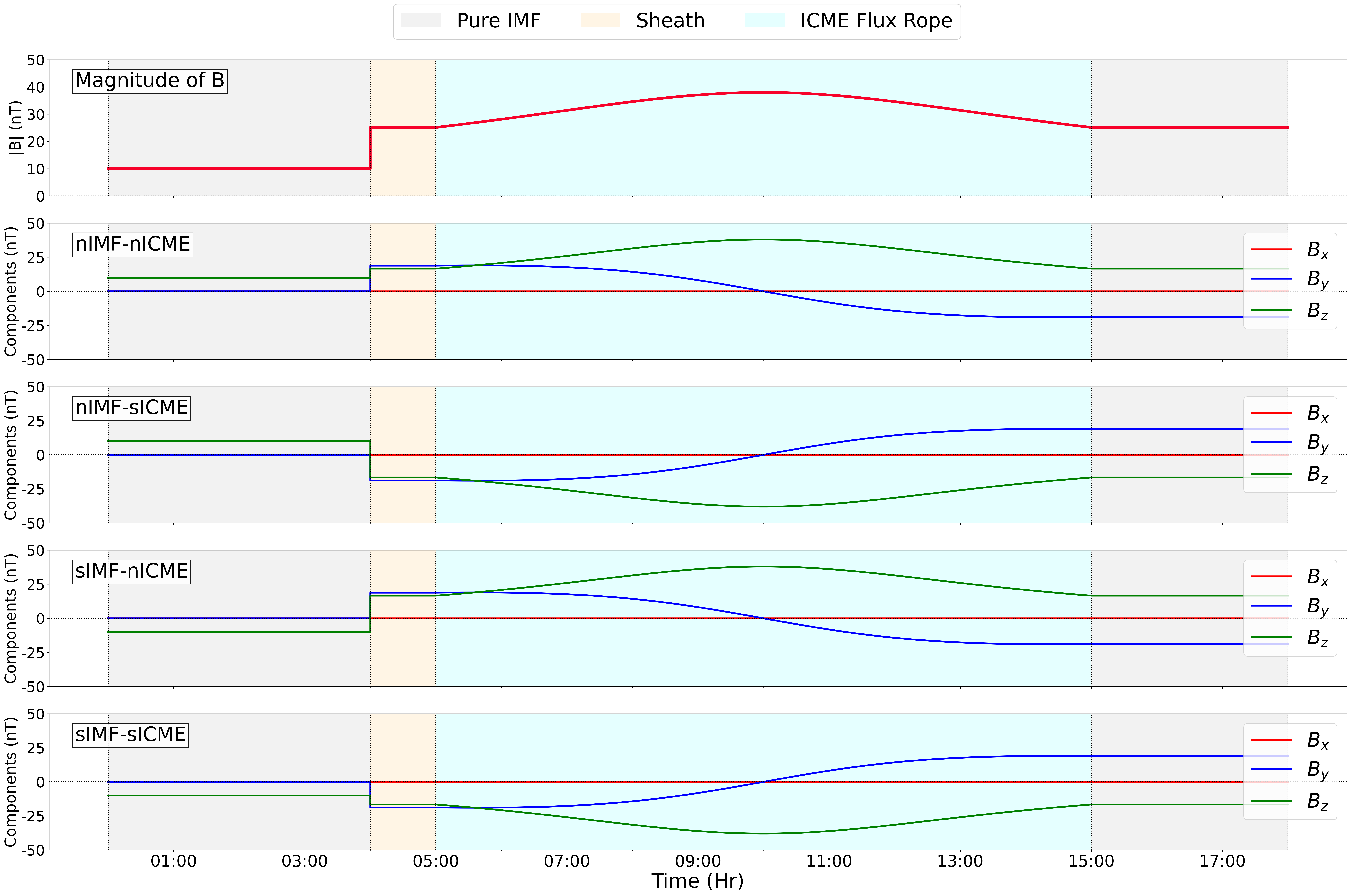}
    \caption{Temporal evolution of the stellar magnetic field conditions used as boundary inputs in the simulations. The top panel shows the total field magnitude ($|B|$), and the lower panels depicts the Cartesian components ($B_x$, $B_y$, $B_z$) for each IMF–ICME orientation case. Shaded regions indicate the pure IMF (grey), ICME sheath (beige), and ICME flux rope (blue) phases followed by pure IMF.}
    \label{IMF_ICME}

\end{figure*}

In this study, we focus on a far-out Earth-like planet with an Earth-like mass, radius and dipolar tilt, while the nominal stellar wind is assumed to have a speed of 350 km/s corresponding to quiet times observed in the Sun-Earth system. The ICME is assumed to have a speed of 1000 km/s. We perform a set of simulations, as illustrated in Fig. \ref{IMF_ICME}, by varying the orientation of the interplanetary magnetic field (IMF) and the embedded ICME magnetic field. Specifically, we consider four cases: northward IMF followed by northward ICME in time (nIMF–nICME), northward IMF followed by southward ICME (nIMF–sICME), southward IMF followed by northward ICME (sIMF-nICME) and southward IMF followed by southward ICME (sIMF–sICME). For each configuration, we also vary the strength of the planetary magnetosphere from unmagnetised to $0.1 {B}_{{e}}, {B}_{{e}}, \thinspace $and$ \thinspace 2 {B}_{{e}}$ and perform simulations with all possible combinations. Here ${B}_{{e}}$ = $3.1\times 10^4$ nT denotes the surface equatorial magnetic field strength of the Earth's dipolar field.  

Fig. \ref{IMF_ICME} presents the temporal evolution of the upstream magnetic field conditions that we consider as boundary inputs in our simulations. The top panel shows the total magnetic field magnitude ($|B|$), while the next four panels display the three Cartesian components ($B_x$, $B_y$, $B_z$) for each IMF–ICME orientation case. The shaded regions distinguish the different phases: the initial steady IMF, the ICME sheath, and the ICME flux rope, followed by a return to the background IMF. In each case, the simulation begins with a constant IMF, after which the ICME sheath arrives, resulting in a step-like increase in $|B|$ as expected in the shock front. The subsequent flux rope phase exhibits a smoothly varying rotation of the field components, representing the passage of the ICME structure. Although the four cases differ in the orientation and temporal evolution of the magnetic field components, we ensure that the total magnetic field magnitude remains the same across all cases, as depicted in the top panel. Our simulation setup enables identification of how field orientation influences planetary magnetospheric responses by maintaining a consistent magnetic pressure across all IMF-ICME cases. 

The input parameters at the stellar wind injection boundary (at $x$= -145 ${R_p}$ in $yz$ plane) are obtained by solving the Rankine-Hugoniot conditions for the given shock velocity. The density and pressure during the ICMEs are also defined by solving the Rankine–Hugoniot conditions to generate and maintain the shock fronts. For all other boundary faces of the Cartesian box, an outflow boundary condition is implemented. We set the magnetic field strength of the ICME by time averaging the corresponding in-situ observations of MESSENGER and Venus Express \citep{S_Good_2016}. The magnetic field orientation of the stellar wind and (or) planetary magnetic field strength are varied in each computational run and the steady state configuration of the planetary magnetosphere, atmospheric loss, global magnetic polarity and magnetotail current density are analyzed. 

\section{RESULTS AND DISCUSSION} \label{sec:floats}

% \subsection{Comparative analysis of atmospheric escape}

The interaction of stellar CMEs and planetary magnetic configurations results in a complex, dynamically evolving system shaping the space environment around the (exo)planetary magnetosphere. Our 3D magnetohydrodynamic (MHD) simulations provide insights into the governing dynamics, and we explore interactions across varying (exo)planetary magnetic field strengths and ICME magnetic field orientations. In subsequent sections, we discuss our analysis focusing on global planetary magnetospheric configurations, magnetic polarity reversal, magnetotail current density evolution, and atmospheric mass loss rates. 

\subsection{Global Magnetic Polarity Reversal Dynamics}

Stellar wind and CMEs play a crucial role in shaping the planetary magnetosphere, with the orientation of the stellar wind magnetic field exerting a significant influence on the configuration of magnetospheric structures. \cite{Vech_2016} discusses the statistical features of global polarity reversal in the Venusian-induced magnetosphere in response to changes in the orientation of the interplanetary magnetic field. We perform 3D MHD simulations to explore the effects of stellar wind and ICME magnetic field orientation on (exo)planetary magnetospheres with different magnetic field strengths. Fig. \ref{pol_rev}(a) illustrate the global magnetic configuration for the unmagnetised and magnetised planet (Bp = 2 Be). The left panels (A $\&$ C) of \ref{pol_rev} (a) show the steady-state configuration when the planetary magnetosphere interacts with only nominal northward oriented stellar wind and the right panels (B $\&$ D) represent the state when planetary magnetosphere interacts with southward oriented ICME. In these plots, yellow and blue lines represent the streamlines of the stellar wind and the planetary magnetic field, respectively. Vectors indicate the direction of the magnetic field, and the current density is plotted as the background in each figure.

% Our simulation for the unmagnetised planet shows a global magnetic polarity reversal of the induced magnetosphere in response to a reversal in the stellar wind magnetic field orientation.

The top panel of Fig. \ref{pol_rev}(a) (A $\&$ B) demonstrates the global polarity reversal for the induced magnetosphere of the unmagnetised planet  which has been conditioned by a northward-oriented stellar wind (thereby picking up a clockwise magnetic field when viewed from positive to negative y direction) followed by a southward-oriented ICME (note the magnetic field becomes anti-clockwise when viewed from the same perspective). As the ICME with oppositely directed magnetic field orientation relative to the nominal stellar wind magnetic field interacts with the planet, magnetic reconnection occurs for a brief duration. Subsequently, the imposed magnetosphere reconfigures and reverses, aligning with the external stellar CME magnetic field orientation. Our analysis indicates that the planetary magnetic field strength influences the susceptibility to stellar wind-induced polarity reversals. For weakly magnetised planets (Bp = 0.1 Be), the stellar wind or a CME magnetic fields can still induce polarity reversals (Fig \ref{pol_rev}(b)). However, as the planetary magnetic field increases (Bp = Be or Bp = 2 Be), the magnetosphere becomes increasingly resistant to reversals even when forced by a CME. It maintains its intrinsic polarity in the vicinity of the planet, when exposed to an oppositely directed ICME (Fig. \ref{pol_rev} (b)). The results suggest that the relative strengths of planetary and stellar wind magnetic fields are fundamental to determining the potential for global polarity reversals in (exo)planetary magnetospheres. A similar reconfiguration of the imposed magnetosphere takes place during the sIMF-nICME event. For nIMF-nICME and sIMF-sICME events, the imposed magnetosphere maintains the same configuration dictated by the unchanging external stellar wind magnetic field orientation. This hevaior is expected and may be considered as the ``control'' experiment. A strongly magnetised planet, maintains its intrinsic polarity for all cases of IMF - ICME events.

Our simulations demonstrate that the response of a planetary magnetosphere to stellar CME-induced polarity reversal is strongly modulated by the planet's intrinsic magnetic field strength. In the absence of an intrinsic magnetic field, the planetary ionosphere interacts directly with the stellar wind, leading to the formation of an induced magnetosphere. The polarity of this induced magnetosphere is dictated by the orientation of the stellar wind magnetic field. When the stellar wind magnetic field reverses its orientation, the induced magnetosphere rapidly reconfigures, resulting in a global polarity reversal. In contrast, planets with strong intrinsic magnetic fields are less susceptible to polarity reversal. The intrinsic field dominates the interaction, shielding the planet from the direct influence of the stellar wind. While, the magnetosphere may compress or undergo local reconnection events, the overall magnetic topology remains stable.

To further illustrate the polarity reversal, we plot the Z-component of the magnetic field at the planetary day side at a distance of 2.8 Rp (Fig \ref{pol_rev}(b)). We select the specific distance because magnetic field measurements closer to the planet consistently reflect the intrinsic polarity (of a magnetised planet) as observed at 2.8 $R_p$, while measurements farther away may represent the stellar wind magnetic field. Fig \ref{pol_rev}(b) shows that at the distance of 2.8 Rp, the magnetic field reverses its direction for unmagnetised and weakly magnetised planets (Bp = 0 and Bp = 0.1 Be) as depicted by the blue and cyan curves, respectively. However, as the planetary magnetic field strength increases ( for cases, Bp = Be, and 2Be), the intrinsic planetary magnetic field (green and red curves) do not flip their direction; i.e. the planet retains its native magnetic field orientation.

\subsection{Magnetotail Current Density Dynamics}

The external stellar wind conditions significantly influence the current density dynamics in the planetary magnetosphere. \cite{Vech_2016} presents observational evidence of diminished heavy ion flux in the near-Venus magnetotail during polarity reversal events -- attributing this phenomenon to the magnetic detachment of the plasma tail from the planetary ionosphere. The drift of charged particles along magnetic field lines governs current density in the magnetotail. A reduction in ion flux impacts the available charge carriers for current flow, leading to a drop in current density. Our MHD simulation approach cannot capture microscale physics, however, qualitatively our results corroborate the observational findings of \cite{Vech_2016}. Our analysis indicates a decrease in current density magnitude near the magnetotail of an unmagnetised planet during polarity reversal events analogous to observations for Venus. 

\begin{figure*}
    \centering
    \includegraphics[width=0.7\textwidth]{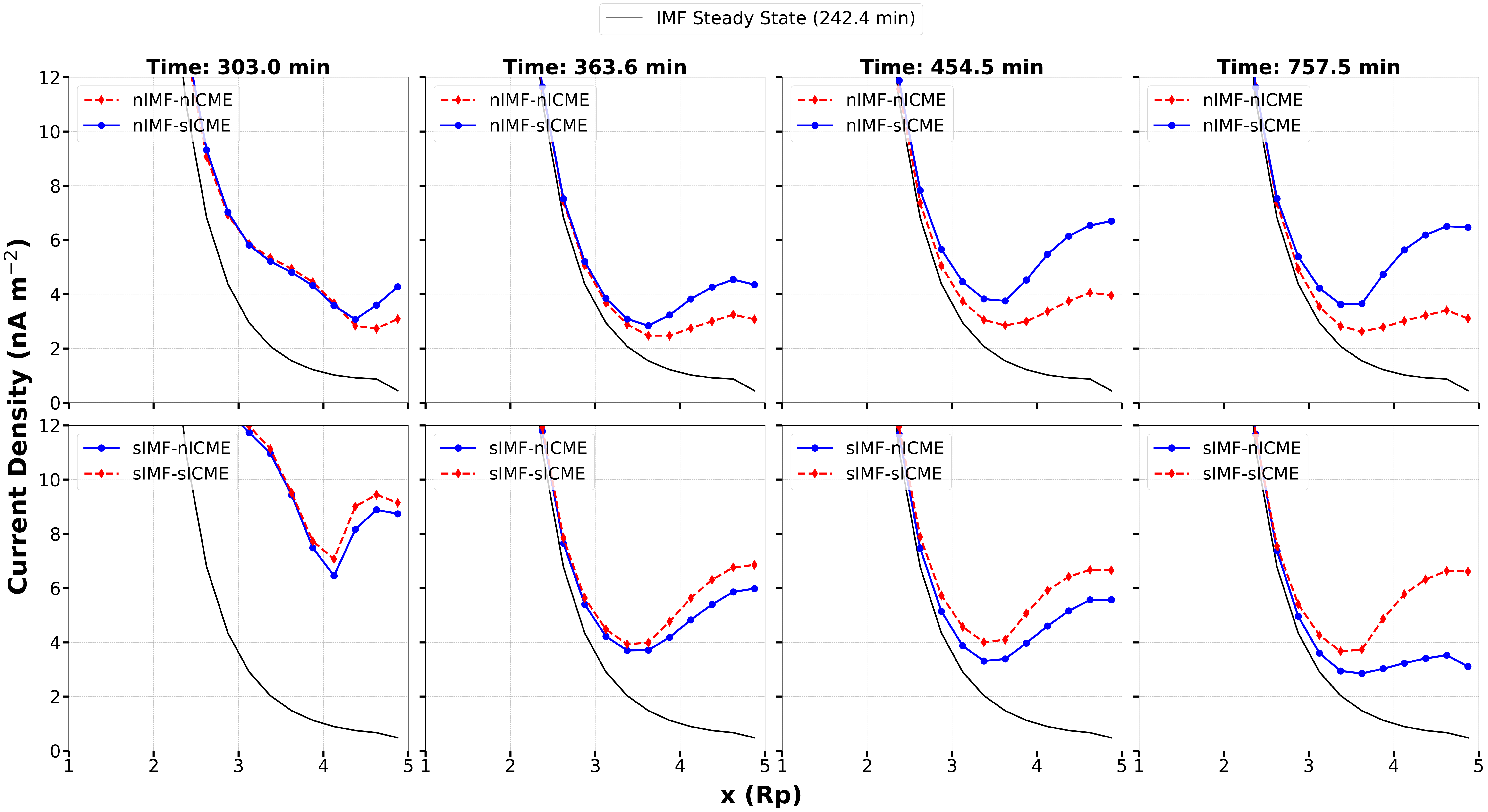}
    \vspace{1mm}
    \par\noindent\textbf{(d) Bp = 2Be}
    \vspace{3mm}
    
    \caption{Current density magnitude evolution in the (exo)planetary magnetotail during ICME interaction. Panels show different (exo)planetary magnetic field strengths: (a) Bp = 0 (unmagnetised), (b) Bp = 0.1Be, (c) Bp = Be, and (d) Bp = 2Be. For each case, the temporal evolution is shown for various ICME magnetic field orientations. The black curve represents steady-state nominal stellar wind interaction, while other curves show different time instances during ICME interaction. The blue curve indicates the scenario where the interplanetary magnetic field (IMF) followed by the ICME field are oppositely directed, whereas the red curve represents the case where the planet is impacted by an IMF followed by the same ICME magnetic field orientation.}
    \label{Jmag}
\end{figure*}

We compute the current density magnitude on the planetary night side (spanning from 1.5 to 5 Rp along the x-axis). To account for current sheet thickness, we average the current density across the yz-plane, ranging from -1 to 1 Rp in both y and z directions. Fig. \ref{Jmag} illustrates the evolution of current density magnitude during ICME interactions for planets with varying magnetic field strengths. Each panel in Fig. \ref{Jmag} (a-d) represents a different planetary magnetic field strength, and within each panel, the current density magnitude is plotted as a function of distance along the x-axis (planetary night side) at different times during the simulation. The black curve in each panel represents the steady-state current density profile during the initial interaction of the planet with nominal stellar wind. Other curves show the current density profiles at other indicates time instances during the ICME impact and passage. The first row in each panel illustrates cases where a northward IMF is following by northward and southward ICMEs, while the second row depicts cases where a southward IMF is followed by northward and southward ICMEs, respectively. For the unmagnetised planet, Fig. \ref{Jmag} (a), we find that the current density is consistently lower in magnitude when an oppositely directed ICME (relative to the previous nominal stellar wind IMF) interacts with the planet. The column-wise plot of Fig. \ref{Jmag} (a) indicates that initially, the current density magnitude is lower for the cases where polarity reversal occurs (1 and 2 columns of Fig. \ref{Jmag}(a)). However, following some time, the current density slowly recovers, regardless of stellar wind magnetic field orientation (column 3 and 4 of Fig. \ref{Jmag}(a)).

For magnetised planets (Fig. \ref{Jmag}b-d) {Bp = 0.1 Be, Be and 2 Be}, current density magnitude is elevated when southward-oriented ICME interacts with the planetary magnetosphere. This elevation can be attributed to enhanced magnetic reconnection between the southward IMF and the northward-oriented planetary magnetic field. Our analysis suggests that this distinct magnetotail current density responses can serve as indicators of a planet's magnetic environment, distinguishing between intrinsic and induced magnetospheres in future observational studies.

\subsection{Atmospheric Escape Rate}

Stellar wind and CMEs are the predominant factor resulting in the atmospheric escape of planets. The rate of atmospheric mass loss depends on the extent to which stellar wind penetrates into the planetary atmosphere. The extent of this stellar wind penetration is largely dependent on magnetic reconnections of the stellar wind magnetic field with the planetary magnetosphere and the dynamic ram pressure. The magnetic field orientations of IMF and ICMEs significantly influence the magnetic reconnection dynamics with the planetary magnetosphere and is therefore expected to influence the atmospheric escape rate. 

We perform a comparative analysis of atmospheric escape rates resulting from the interactions of different stellar wind and CME magnetic field orientations relative to the planetary magnetosphere. To compute the total mass-loss rate, we consider a cube of edge length 10 Rp (extending from –5 Rp to 5 Rp in all three directions) with its geometric center coinciding with the centre of the planet such that the whole planet -- including its modeled atmosphere -- is enclosed within the cube. Mass-loss rates are computed across all six faces of the cube and loss rates from all six faces are summed up to obtain the total mass-loss rate. In our study, atmospheric escape is primarily driven by the magnetic reconnection dynamics between the stellar forcing and the planetary magnetosphere. Atmospheric losses can be caused by stellar winds, coronal mass ejections (CMEs), or high-energy radiation. However, we limit our focus here on losses driven by magnetised CMEs and stellar winds, rather than those caused by high energy radiation or ion pickup. These processes could further enhance the atmospheric escape (see, e.g., \cite{Hazra_2025}).

\begin{figure*}
    \centering
    \includegraphics[width=0.9\textwidth]{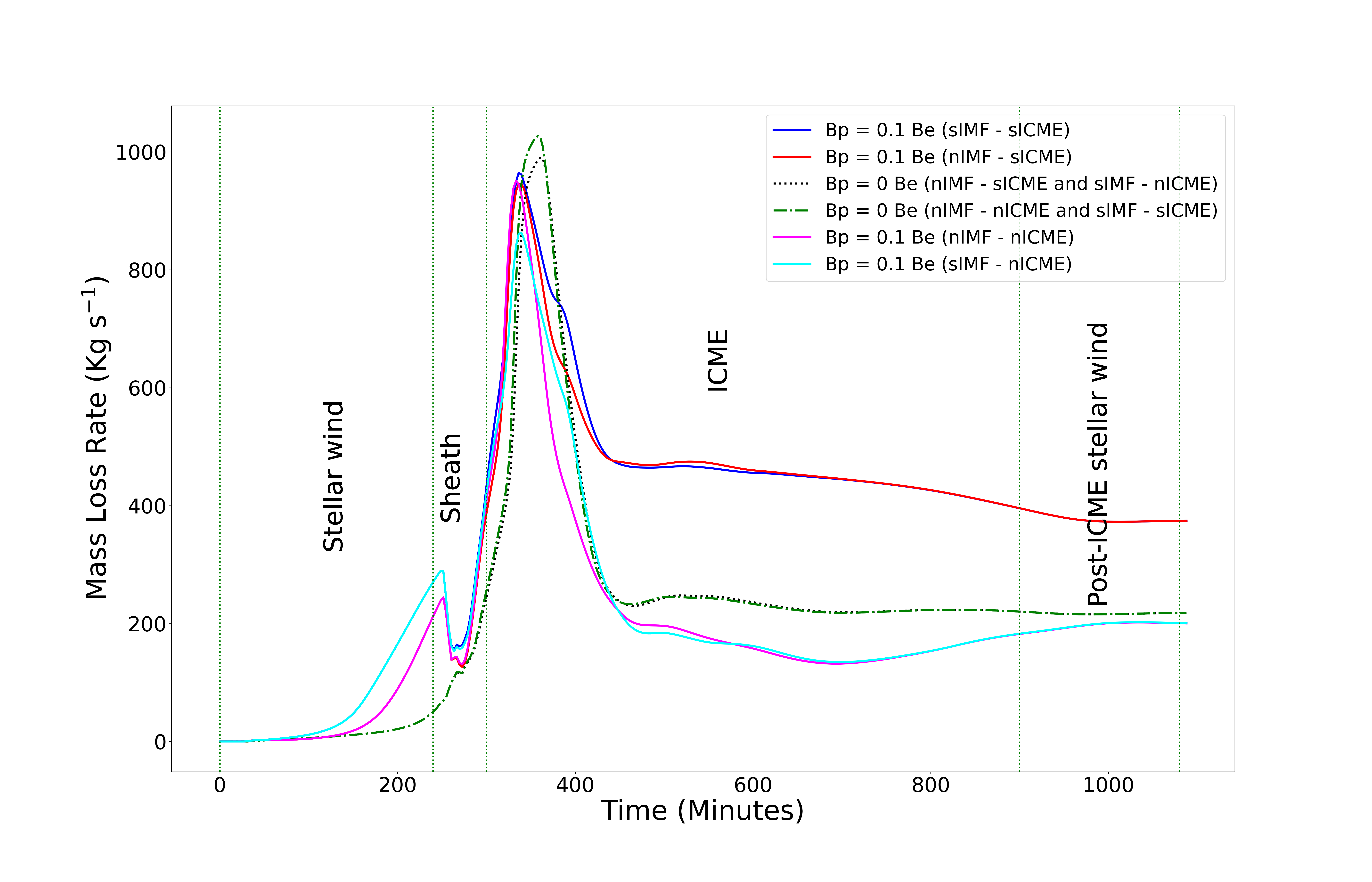}
    \caption{Temporal evolution of atmospheric mass-loss rates for unmagnetised and weakly magnetised (Bp = 0.1Be) (exo)planets during stellar field interaction. Mass-loss rates are computed within a cubic volume (± 5 Rp in all directions) encompassing the planet and its atmosphere. Different curves represent various IMF-ICME orientation combinations, with vertical dashed lines marking key transition points: stellar wind, sheath region, ICME impact, and post- ICME stellar wind conditions.}
    \label{atm_loss}
\end{figure*}

Fig. \ref{atm_loss} illustrates the temporal evolution of atmospheric mass-loss rates for unmagnetised and weakly magnetised (Bp = 0.1Be) planets during ICME interaction. Different plots show that for all cases of our simulations, the mass loss rate initially experiences a brief dip as the ICME impacts the planet. This dip occurs due to the compression and confinement of the magnetosphere and modeled atmosphere caused by the impact of the ICME shock. However, following this compression, the mass loss rate rises sharply as a result of MHD interactions between the ICME plasma and the modeled planetary atmosphere, mediated via magnetic reconnection and viscous coupling. Eventually, the mass loss rate reduces and stabilizes slowly in the recovery phase from the ICME impact.

For all four combinations of IMF and ICME magnetic-field orientations for unmagnetised planets, the atmospheric mass loss rates are nearly similar. In contrast, for a magnetised planet (Bp = 0.1 Be), the mass loss rates are significantly influenced by the orientation of the stellar magnetic field. Southward ICMEs induce much higher mass loss rates compared to northward ICMEs in the case of magnetised planets (with a negative, Earth-like dipole field). For cases when the orientation of the stellar wind magnetic field aligns with the planetary magnetic field (i.e. both are northward oriented), the presence of a planetary magnetic field can provide shielding against atmospheric mass loss compared to the unmagnetised planet. 

Unlike the case of a magnetised planet, in unmagnetised planets reconnection dynamics does not play a major role in induced magnetospheres, and thus, the escape rate is nearly similar for different orientations of the magnetic field in unmagnetised planets. In Fig. \ref{atm_loss}, we show a comparison of atmospheric mass loss rates for Bp = 0.1 Be and an unmagnetised planet. We consider only these cases because as we increase the planetary magnetic field, the box size to compute the atmospheric mass loss rate also needs to be larger. Otherwise, the magnetopause boundary crosses the boundary of the box during the temporal evolution resulting in spurious results. We have verified that for the other two cases as well (i.e. Bp = Be and 2Be), our results are qualitatively similar to those for Bp = 0.1Be but with a larger box size. For the plasma atmosphere, as considered in the present study, the mass loss rate is higher for the planet with a stronger planetary magnetic field. This shows that intrinsic magnetic fields do not always protect the planet against mass loss and the behaviour can be quite complex depending on diverse scenarios \citep{Egan, refId0, Gupta_2023}.

\section{CONCLUSIONS} \label{sec:displaymath}

In this study, we employ 3D global MHD modelling to investigate how varying orientations of stellar wind and ICME magnetic fields impact (exo)planetary environments. We explore the impact of varying orientations of stellar wind magnetic field on atmospheric escape rates, global magnetospheric polarity dynamics and magnetotail current density evolution across planets with varying magnetic field strengths, ranging from unmagnetised to magnetised planets. We focus only on far-out Earth-like planetary systems.

For unmagnetised planets, we find that the mass loss rate does not strongly depend on the stellar wind and stellar CME magnetic field orientation. As an ICME impacts the planet there is a compression in the planetary magnetosphere leading to a momentarily dip in the atmospheric mass loss rates. After the compression, the ICME passage leads to substantially higher atmospheric escape rates than nominal stellar wind conditions. In the case of a magnetised planet, when the planetary and stellar wind magnetic fields are aligned in the same direction, the planetary magnetosphere provides shielding against atmospheric escape compared to an unmagnetised planet, which is expected. Intrinsic planetary magnetic fields, however, do not always protect the planet against atmospheric escape. It can lead to enhanced mass loss in some cases depending upon the orientation of the stellar wind magnetic field. 

Most notably, our simulations show that unmagnetised planets undergo global magnetic polarity reversals when the stellar wind magnetic field or stellar CME orientation reverses relative to the preexisting ambient interplanetary magnetic field conditions. This polarity reversal phenomena becomes less pronounced with increasing planetary magnetic field strength. These theoretical findings are supported by previous observational studies -- such as those gleaned from Venus -- which exhibit global polarity reversals under similar conditions. 

We analyze the current density magnitude near the planet at the magnetotail side. Our results show that during polarity reversals, the current density magnitude temporarily drops for the case of an unmagnetised planet. For the magnetised planet, the current density magnitude is higher for impacts of ICMEs with opposite magnetic field orientation with respect to the planetary magnetic field. This is likely due to reconnection dynamics dominating the star-planet interactions in this case. We speculate that the observation of a drop in heavy ion flux in the magnetotail during polarity reversals in the Venusian environment \cite{Vech_2016} is due to this drop in current density (which is manifest in our simulations for an unmagnetised planet).

Our findings have important implications for distinguishing between unmagnetised and magnetised planets and for devising future observational strategies for constraining (exo)planetary magnetism. Furthermore, insights on atmospheric dynamics forced by stellar plasma winds and magnetic storms gleaned from our simulations have relevance for characterizing habitability of exoplanets from the perspective of star-planet interactions mediated via magnetic fields.

\section*{acknowledgments}

The development of the Storm Interaction module (CESSI-STORMI) and the simulations were carried out at the Center of Excellence in Space Sciences India (CESSI), supported by IISER Kolkata, Ministry of Education, Government of India. SG acknowledges Dr. Jagannath Nayak, Director CHESS, and Saumendra Nath Datta, Technology Director at CHESS, for the necessary approval and encouragement to continue her PhD thesis research work at CESSI. Authors extend their gratitude to Dr. Arnab Basak for insightful discussions.

\section*{DATA AVAILABILITY}

All simulation runs in this study are performed using the STORM Interaction Module (CESSI-STORMI), that is developed based on the open-source 3D MHD PLUTO code. Data from our simulations will be shared upon reasonable request to the corresponding author.



\bibliographystyle{mnras}
\bibliography{ICME_planet} % if your bibtex file is called example.bib

%%%%%%%%%%%%%%%%%%%%%%%%%%%%%%%%%%%%%%%%%%%%%%%%%%

% First part: Figures (a) and (b)

\begin{figure*}
    \centering
    \includegraphics[width=\textwidth]{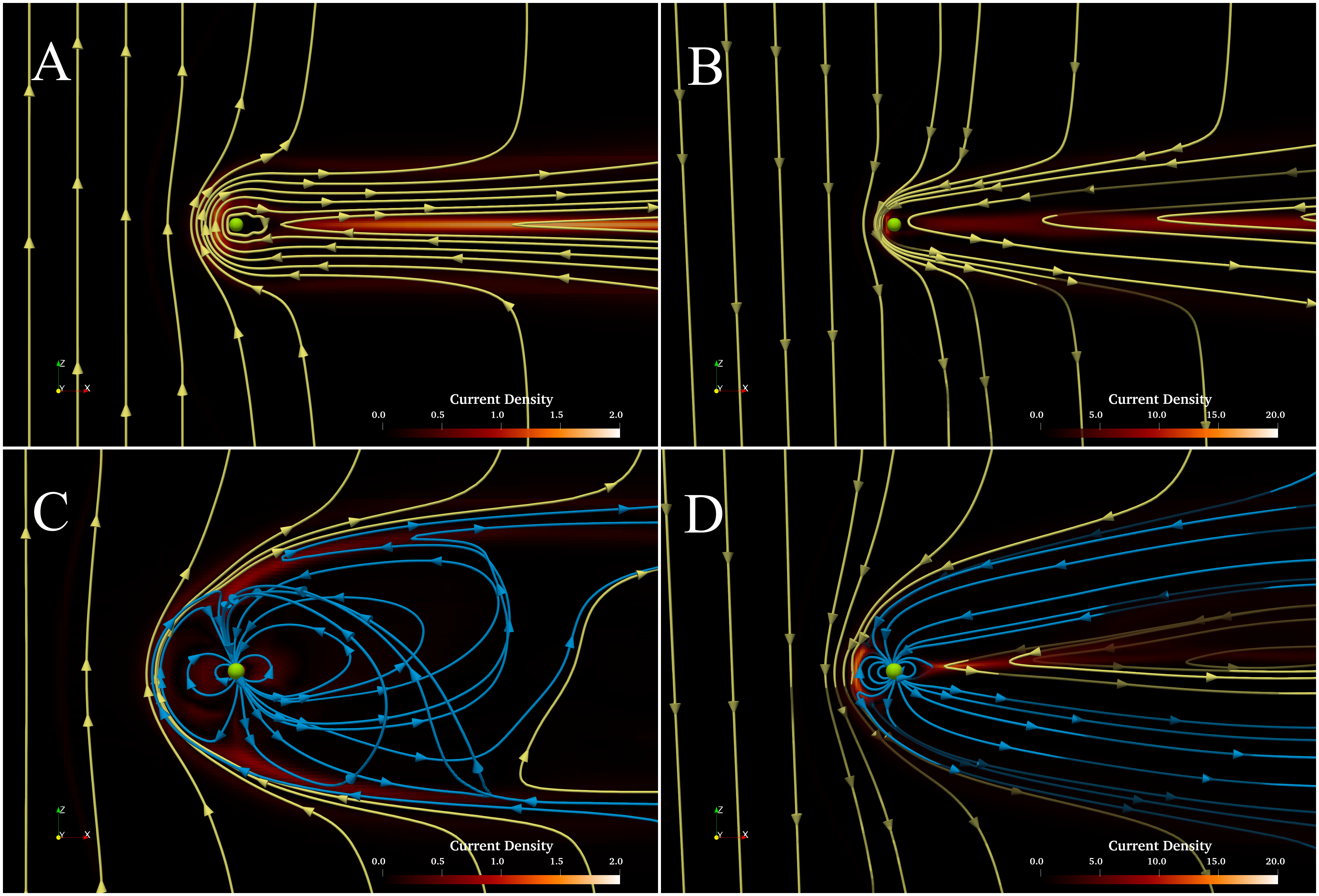}
    \vspace{1mm}
    \par\noindent\textbf{(a)}
    \vspace{4mm}
    
    \includegraphics[width=0.8\textwidth]{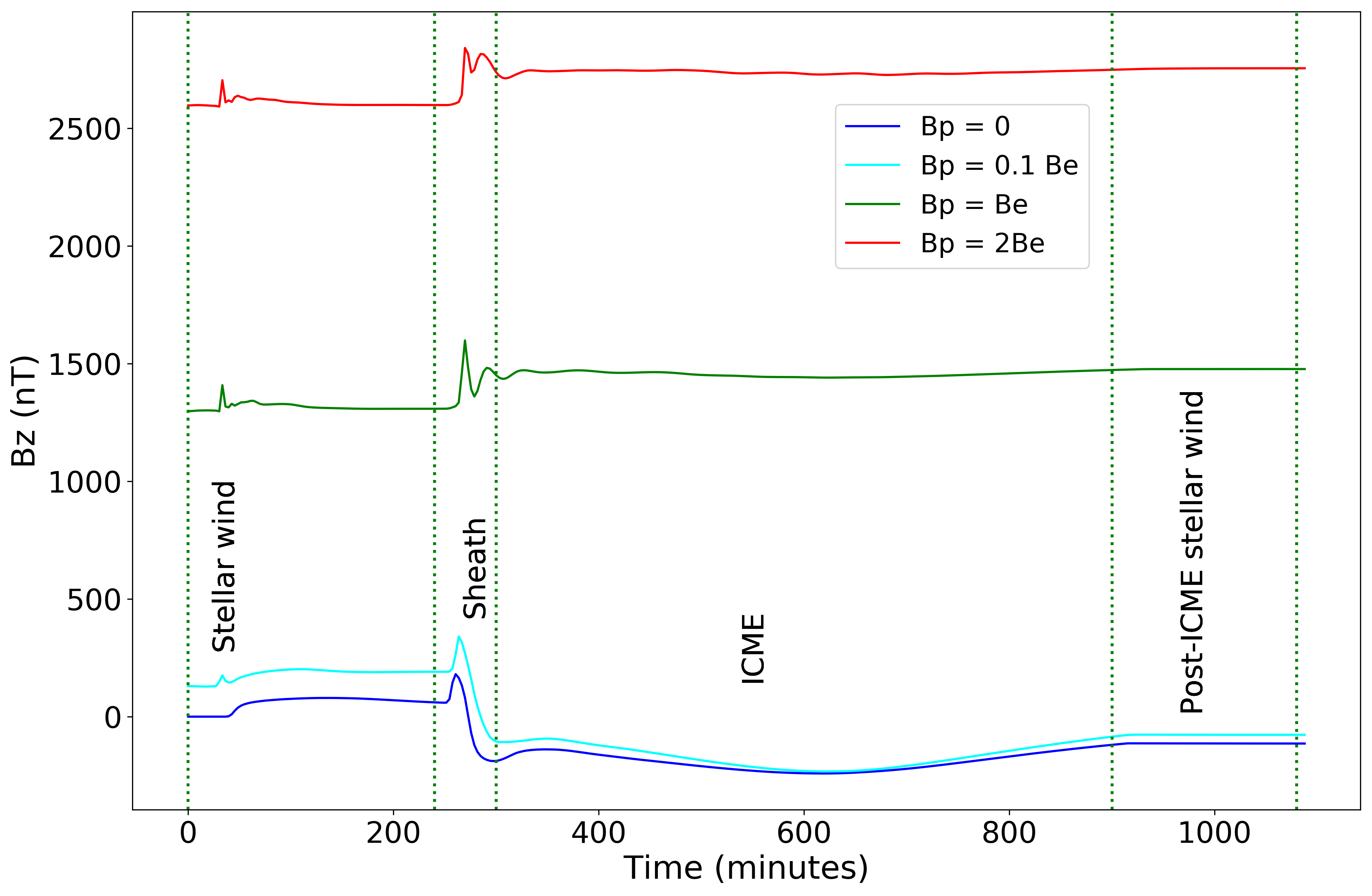}
    \vspace{1mm}
    \par\noindent\textbf{(b)}
\end{figure*}

% Second part: Caption on next page
\begin{figure*}
    \centering
    \vspace*{1.5cm} % Adjust to minimize gap
    \caption{Polarity reversal in (exo)planetary magnetospheres. (a) Evolution of magnetic field streamlines, blue representing (exo)planetary magnetic field while yellow represent stellar wind magnetic field lines during planetary interaction with the stellar wind. (A) is the steady state configuration when northward stellar wind interacts with the unmagnetised planet (t = 121.6 min), (B) shows the magnetospheric configuration of unmagnetised planet during the southward-oriented ICME interaction with the planet (t = 317.68 min). The plots show the global polarity reversal when transitioning from northward nominal stellar wind to southward ICME. (C) and (D) shows similar interaction for a magnetised planet (Bp = 2Be), showing that polarity does not reverse in the vicinity of the planet. (b) Z-component of the magnetic field at 2.8 Rp on the planetary dayside, illustrating that the polarity reversal at the vicinity of the planet diminishes with increasing (exo)planetary magnetic field strength.}
    \label{pol_rev}
\end{figure*}

% First page: 3 panels
\begin{figure*}
    \centering
    \includegraphics[width=0.7\textwidth]{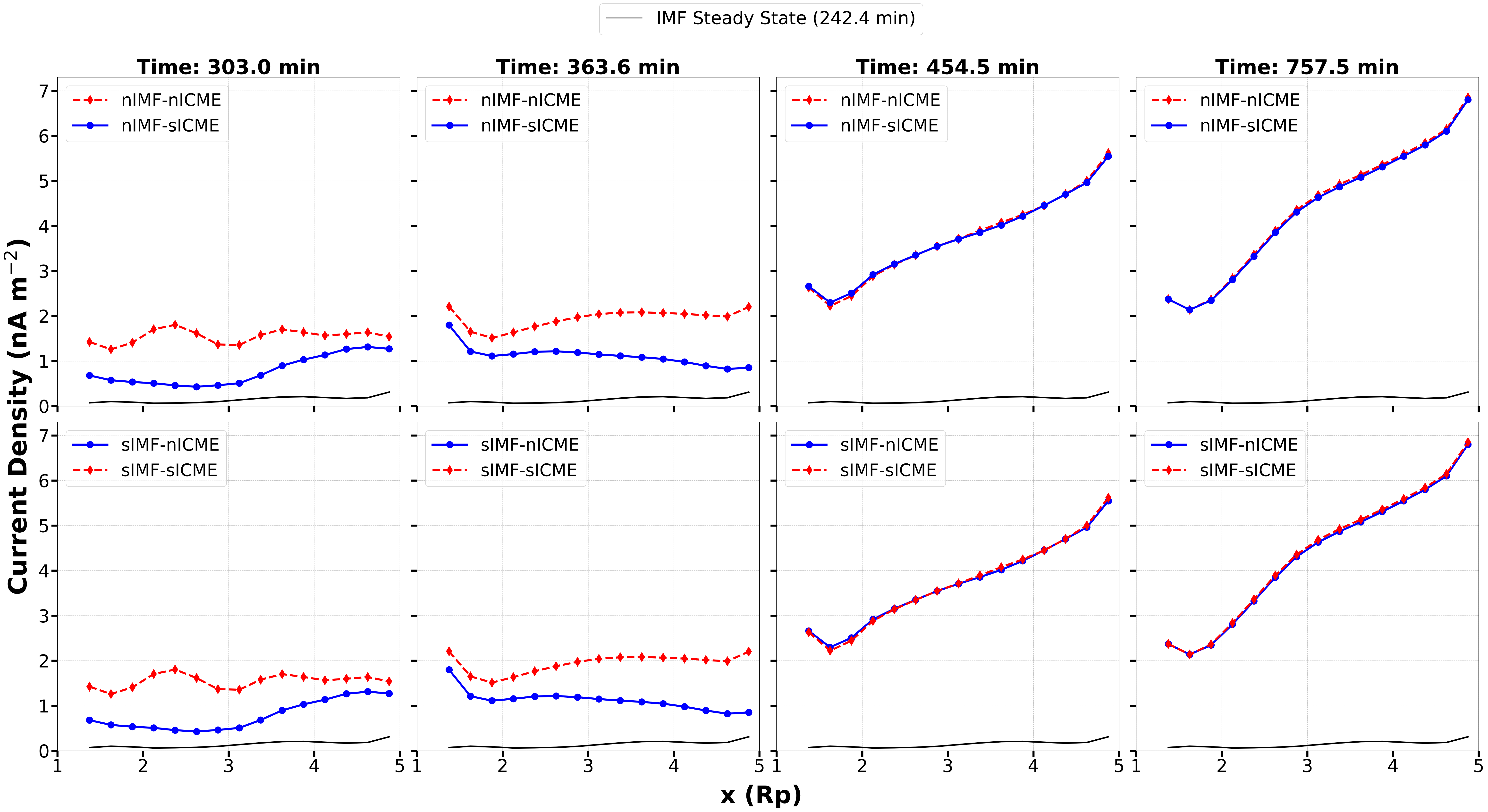}
    \vspace{1mm}
    \par\noindent\textbf{(a) Bp = 0}
    \vspace{3mm}
    
    \includegraphics[width=0.7\textwidth]{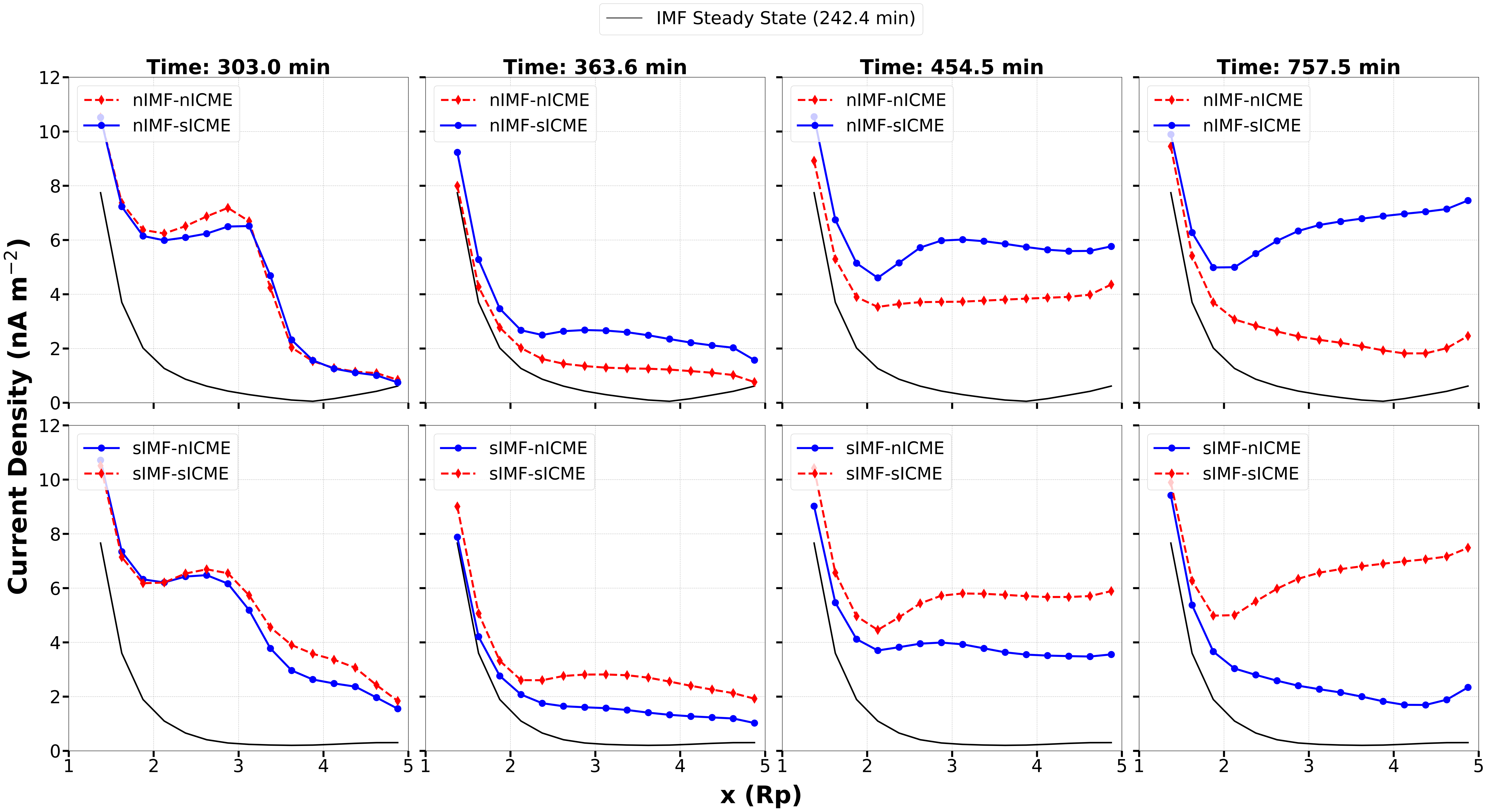}
    \vspace{1mm}
    \par\noindent\textbf{(b) Bp = 0.1Be}
    \vspace{3mm}
    
    \includegraphics[width=0.7\textwidth]{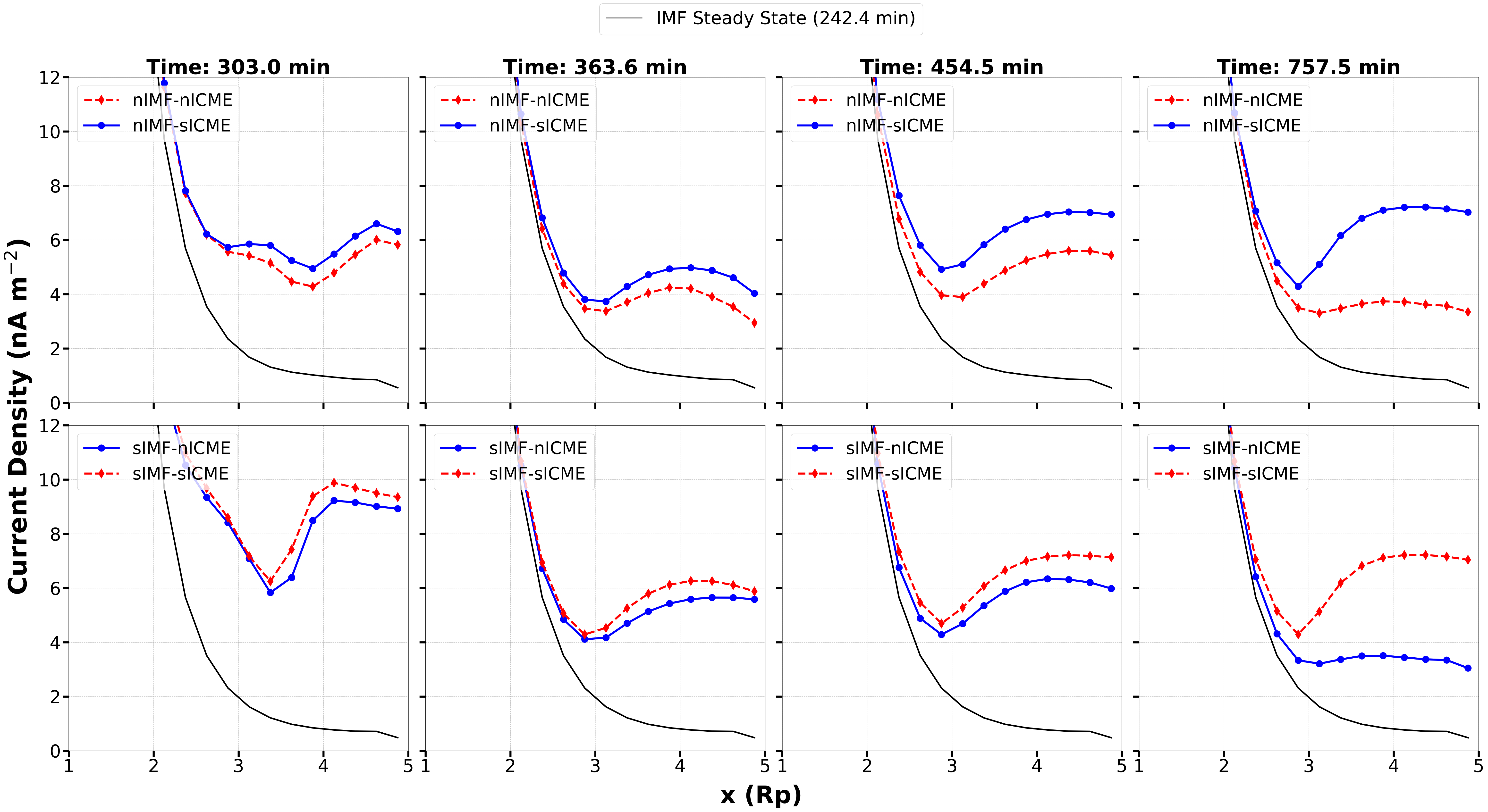}
    \vspace{1mm}
    \par\noindent\textbf{(c) Bp = Be}
\end{figure*}

% Second page: 4th panel + caption

% Don't change these lines
% \bsp	% typesetting comment
\label{lastpage}
\end{document}